\documentclass{book}
\usepackage[latin9]{inputenc}
\usepackage{units}
\usepackage{amsmath}
\usepackage{graphicx}

\makeatletter
\usepackage{piers}
\usepackage{cite}
\renewcommand{\citedash}{\citeright\hbox{--}\penalty\@m\citeleft}
\renewcommand{\citepunct}{\citeright,\penalty\@m\hskip.13emplus.1emminus.1em\citeleft}

\usepackage[printonlyused]{acronym}
\acrodef{MAP}{maximum a posteriori probability}
\acrodef{ML}{maximum likelihood}
\acrodef{QAM}{quadrature amplitude modulation}
\acrodef{QPSK}{quadrature phase-shift keying}
\acrodef{ISI}{intersymbol interference}
\acrodef{GLME}{Gelfand-Levitan-Marchenko equation}
\acrodef{NFDM}{nonlinear frequency-division multiplexing}
\acrodef{NFT}{nonlinear Fourier transform}
\acrodef{FNFT}{forward NFT}
\acrodef{BNFT}{backward NFT}
\acrodef{DF-BNFT}{decision-feedback BNFT}
\acrodef{DF-FNFT}{decision-feedback FNFT}
\acrodef{I-FNFT}{incremental FNFT}
\acrodef{OFDM}{orthogonal frequency-division multiplexing}
\acrodef{TX}{transmitter}
\acrodef{RX}{receiver}
\acrodef{FT}{Fourier transform}
\acrodef{DAC}{digital-to-analog converter}
\acrodef{ADC}{analog-to-digital converter}
\acrodef{LP}{Layer-Peeling}
\acrodef{SNR}{signal-to-noise ratio}
\acrodef{NIS}{nonlinear inverse synthesis}
\acrodef{DBP}{digital backpropagation}
\acrodef{QAM}{quadrature amplitude modulation}
\acrodef{SNR}{signal to noise ratio}
\acrodef{SER}{symbol error rate}
\acrodef{SE}{spectral efficiency}
\acrodef{EDC}{electronic dispersion compensation}
\acrodef{NLSE}{nonlinear Schr\"odinger equation}
\acrodef{AWGN}{additive white gaussian noise}
\acrodef{GVD}{group velocity dispersion}

\makeatother

\begin{document}

\title{Improved Detection Strategies for\\
Nonlinear Frequency-Division Multiplexing}

\maketitle

\author{S. Civelli}

\affiliation{TeCIP Institute, Scuola Superiore Sant'Anna}

\address{Via G. Moruzzi, 1}

\city{Pisa}

\postalcode{56124}

\country{Italy}

\phone{}

\fax{}

\email{s.civelli@santannapisa.it}

\misc{}

\nomakeauthor{}

\author{E. Forestieri}

\affiliation{TeCIP Institute, Scuola Superiore Sant'Anna}

\address{Via G. Moruzzi, 1}

\city{Pisa}

\postalcode{56124}

\country{Italy}

\phone{}

\fax{}

\email{forestieri@sssup.it}

\misc{}

\nomakeauthor{}

\author{M. Secondini}

\affiliation{TeCIP Institute, Scuola Superiore Sant'Anna}

\address{Via G. Moruzzi, 1}

\city{Pisa}

\postalcode{56124}

\country{Italy}

\phone{}

\fax{}

\email{marco.secondini@sssup.it}

\misc{}

\nomakeauthor{}
\begin{authors}
\textbf{S. Civelli}$^{1}$ \textbf{, E. Forestieri}$^{1}$ \textbf{and
M. Secondini}$^{1}$\\
\medskip{}
$^{1}$TeCIP Institute, Scuola Superiore Sant'Anna, Italy\\
\end{authors}

\begin{paper}
\begin{piersabstract}
Two novel detection strategies for nonlinear Fourier transform-based
transmission schemes are proposed. We show, through numerical simulations,
that both strategies achieve a good performance improvement (up to
$\unit[3]{dB}$ and $\unit[5]{dB}$) with respect to conventional
detection, respectively without or only moderately  increasing the
computational complexity of the receiver.
\end{piersabstract}

\psection{Introduction}

Current optical fiber communication systems are limited by the Kerr
nonlinearity, which is one of the main impairments hindering the increase
of the transmission rate. To overcome this limitation and master nonlinearity,
in the recent years, nonlinear spectrum modulation paradigms \cite{Hasegawa93,turitsyn2017optica,Yousefi2014_NFT,Yousefi2016,kamalian2016periodic,le2014nonlinear,le2017nature,civelli2017noise}
have been investigated, devising the integrability of the \ac{NLSE}\textemdash which
models the propagation of a signal in a optical fiber\textemdash with
the \ac{NFT} \cite{ZakSha72,Yousefi2014_NFT,turitsyn2017optica}.
The \ac{NFT} is a sort of nonlinear analogue of the standard linear
\ac{FT}, defining a nonlinear spectrum that undergoes just a phase
rotation during propagation along the optical channel. The umbrella
term \ac{NFDM} indicates \ac{NFT}-based transmission schemes that
encode the information directly on the nonlinear spectrum, such that
deterministic propagation effects\textemdash dispersion and nonlinearity\textemdash can
be exactly removed at the \ac{RX} with a single-tap operation. However,
despite its theoretical robustness against nonlinearity, it is not
yet clear whether \ac{NFDM} can outperform conventional systems \cite{civelli2017noise,Yousefi2016}.
Nevertheless, research about \ac{NFDM} is still in progress, and
\ac{NFDM} schemes are far away from being fully optimized.

\ac{NFDM} paradigms have been developed borrowing concepts from linear
communication and, indeed, can be thought as a nonlinear version
of the well known \ac{OFDM}, using a \ac{BNFT} at the \ac{TX} to
encode information, and a \ac{FNFT} at the \ac{RX} to recover it.
 The detection strategy commonly considered for \ac{NFDM}, also
borrowed from linear systems and optimal for an \ac{AWGN} channel,
is not optimal for \ac{NFDM} since it does not account for the actual
statistics of noise in the nonlinear frequency domain. Therefore,
the currently achieved \ac{NFDM} performance can be much improved,
and novel detection strategies tailored for \ac{NFDM} might reveal
its potential and allow to actually outperform conventional systems.
A first attempt towards this direction is represented by the \ac{DF-BNFT}
detection strategy \cite{civelli2018OFC,civelli2018OPEX}, which provides
a significant performance improvement with respect to conventional
\ac{NFDM}, at the expense of a significant computational complexity.
In this paper, we introduce two novel detection strategies for \ac{NFDM}
by exploting the same causality property of the \ac{NFT} on which
\ac{DF-BNFT} detection is based. We compare this two novel strategies
with standard detection and with \ac{DF-BNFT}, both in terms of performance
and computational complexity.

\psection{System description and detection strategies }

The system setup is sketched in Fig.\ \ref{fig:NFDM}. The \ac{TX},
similarly to the \ac{NIS} technique \cite{le2014nonlinear}, modulates
a train of pulses $g(t)$ with $N_{b}$ \ac{QPSK} symbols, and its
\ac{FT} is mapped onto the continuous part of the nonlinear spectrum
$\rho(\lambda)$. Deterministic propagation effects are removed (full
precompensation) multiplying the nonlinear spectrum by $\exp(4j\lambda^{2}L)$,
$L$ being the normalized channel length, and a \ac{BNFT} is performed
to obtain the samples of $q'(t)$. The optical signal is then obtained
with a \ac{DAC} as $q(t)=q'(-t)$ and launched into the fiber. Further
details about the \ac{TX} setup can be found in \cite{civelli2018OPEX}.
At the \ac{RX}, the samples of the received noisy signal $\tilde{r}(t)$
are obtained with an \ac{ADC} and used for detection. In this work,
we consider the four different detection strategies schematically
depicted in Fig.\ \ref{fig:scheme_rx} and described later in this
section. The \ac{I-FNFT} and \ac{DF-FNFT} strategies are introduced
in this paper for the first time.

\begin{figure}
\includegraphics[width=1\textwidth]{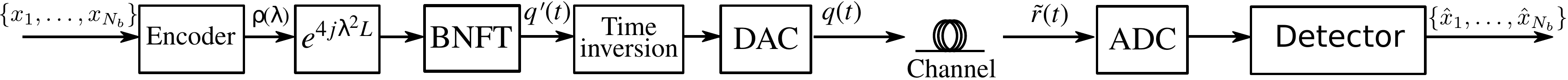}\caption{\label{fig:NFDM}NFDM transmission scheme. For detector details see
Fig.\ \ref{fig:scheme_rx}.}
\end{figure}

\begin{figure}
\includegraphics[width=0.5\textwidth]{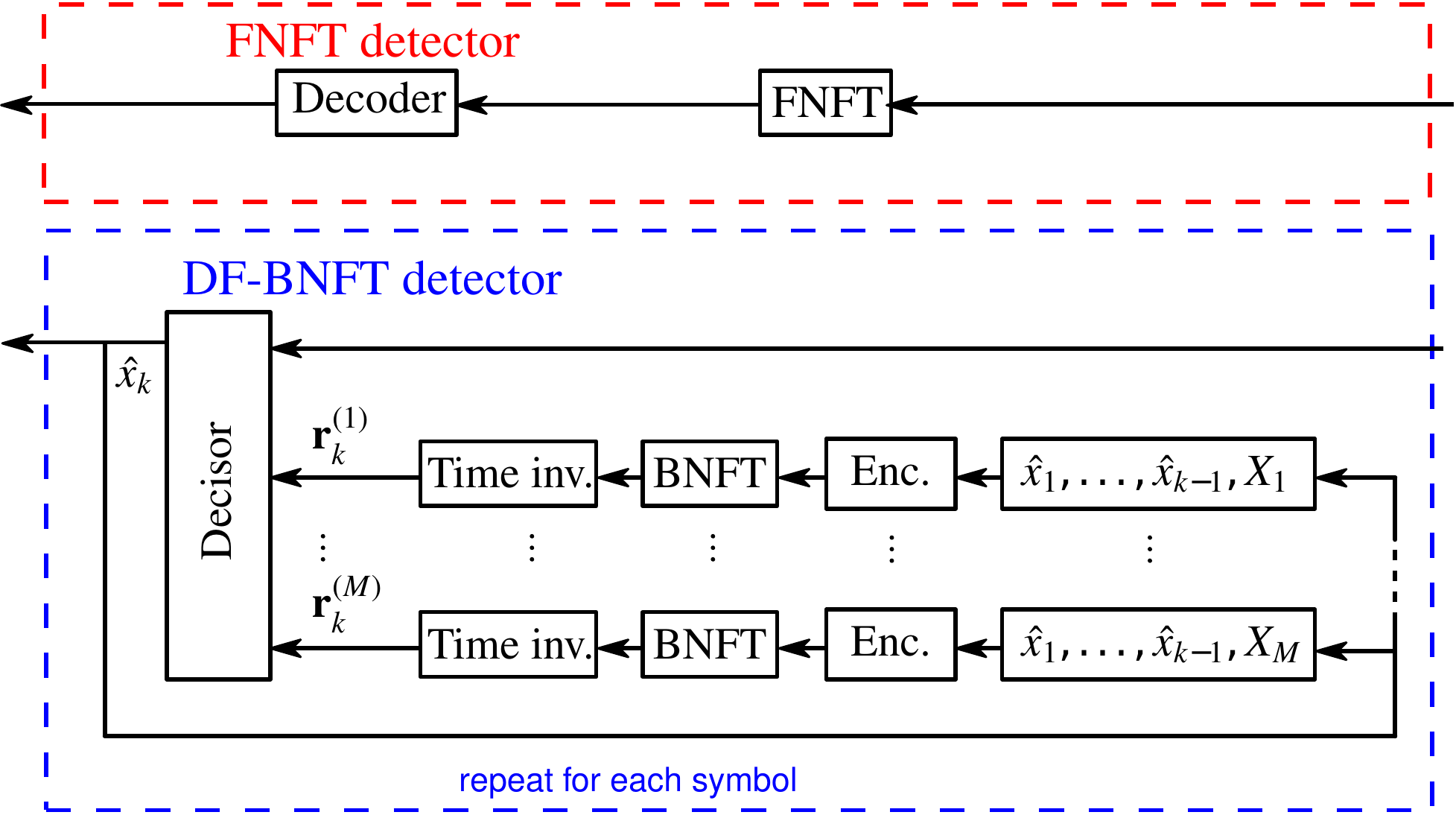}\includegraphics[width=0.5\textwidth]{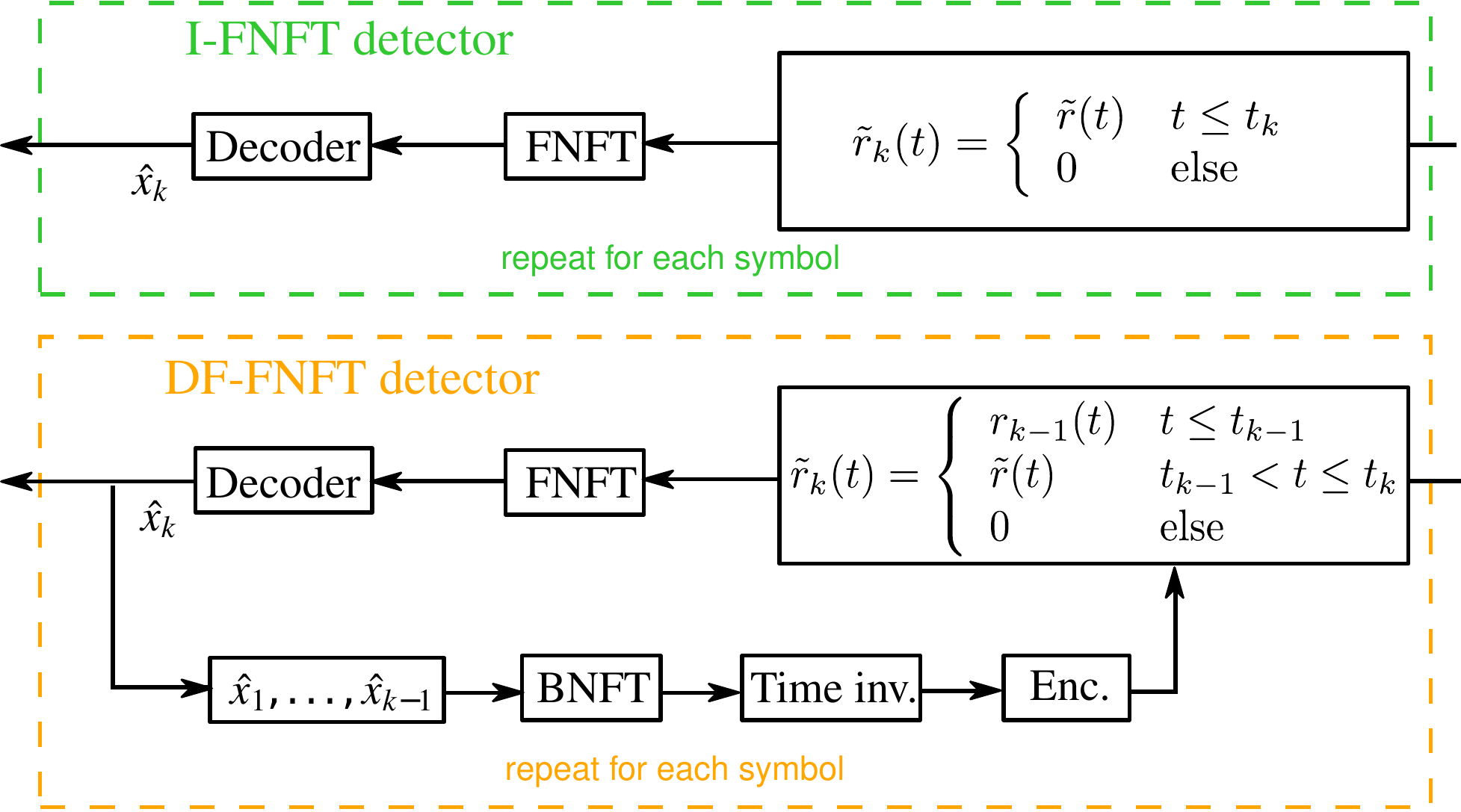}\caption{\label{fig:scheme_rx}Decoders of different detection strategies for
NFDM: FNFT (up left), DF-BNFT (down left), I-FNFT (up right), and
DF-FNFT (down right).}
\end{figure}

An important causality property of the \ac{NFT} applies to the considered
transmission scheme as follows. Let $T_{s}$ be the symbol time, and
$t_{k}=(k-1/2)T_{s}$, such that $(t_{k-1},t_{k}]$ is the symbol
time corresponding to the $k$-th symbol. Let $r(t)$ be the optical
signal obtained propagating $q(t)$ in the ideal noiseless channel,
i.e., the optical signal obtained at the \ac{TX} if precompensation
is not employed. If the pulsewidth is shorter than the symbol time
$T_{s}$, then $r(t)$ for $t\le t_{k}$ depends only the first $k$
symbols $x_{1},\dots,x_{k}$, and not on the next ones $x_{k+1},\dots,x_{N_{b}}$
\cite{civelli2018OPEX}. This causality property can be employed,
as explained in the following, to detect symbols in an iterative way,
and using decision feedback.

The more conventional detection strategy for \ac{NFDM}\textemdash referred
to as \ac{FNFT} detection in the following\textemdash consists in
computing the \ac{FNFT} of the received signal to obtain the nonlinear
spectrum,  and then detecting symbols in the nonlinear frequency
domain after matched filtering and symbol-time sampling \cite{le2014nonlinear,turitsyn2017optica,Turitsyn_nature16,Yousefi2014_NFT,tavakkolnia2016,civelli2018OPEX}.
This simple detection strategy would be optimal if the noise in the
nonlinear spectrum were \ac{AWGN}, which, however, is true only at
low power. Indeed, at higher power, signal and noise interact during
propagation (much like in conventional systems) and, more importantly,
the \ac{FNFT} operation\textemdash a nonlinear operation\textemdash significantly
affects noise statistics. More details about this can be found in
\cite{civelli2018OPEX}. The \ac{DF-BNFT} detection strategy, already
introduced and investigated in previous papers \cite{civelli2018OPEX,civelli2018OFC},
avoids this problem by detecting symbols in time-domain (before the
\ac{FNFT}), using decision feedback and \ac{BNFT}. This detection
strategy, though not optimal, provides a significant performance improvement
(up to $7$dB \cite{civelli2018OPEX}), but requires to compute $M$
\ac{BNFT}s, $M$ being the constellation order. A detailed investigation
about \ac{DF-BNFT} is available in \cite{civelli2018OPEX}. 

The \ac{I-FNFT} strategy is based on deciding symbols in an iterative
way in the nonlinear frequency domain, analyzing only a portion of
the received signal to reduce noise in the nonlinear spectrum, which
increases with signal energy \cite{Turitsyn_nature16}. Specifically,
the $k$-th symbol is detected operating as in conventional \ac{FNFT}
detection (i.e., \ac{FNFT}, matched filtering, and sampling), but
on the signal
\begin{equation}
\tilde{r}_{k}(t)=\begin{cases}
\tilde{r}(t) & t\leq t_{k}\\
0 & \text{else}
\end{cases}.
\end{equation}
This detection strategy can be implemented with the same computational
complexity of \ac{FNFT} detection. Indeed, the nonlinear spectrum
is computed (e.g., with the Boffetta Osborne method \cite{Yousefi2014_NFT,turitsyn2017optica})
recursively adding a small portion of the optical signal and multiplying
for the transfer matrix of this contribution; in our case, this means
that at the $k$-th step, one has already computed the contribution
of the optical signal for $t\leq t_{k-1}$ and needs to add only the
contribution of the signal in $(t_{k-1},t_{k}]$, resulting overall
in a single \ac{FNFT}.

The \ac{DF-FNFT} strategy adds a further step to the \ac{I-FNFT}
one: besides considering only a portion of the signal in detection,
it also takes advantage of the feedback given by already decided symbols
to \emph{clean} the received signal. Specifically, given the symbols
$\hat{x}_{1},\dots,\hat{x}_{k-1}$ already decided, the $k$-th symbol
is decided with two steps: (i) digitally perform a \ac{BNFT} to obtain
for $t<t_{k-1}$ the noiseless signal $r_{k-1}(t)$ which corresponds
to the symbol sequence $\hat{x}_{1},\dots,\hat{x}_{k-1}$ (this is
obtained performing the same operation of the \ac{TX}, but for precompensation),
and (ii) perform standard detection (i.e., \ac{FNFT}, matched filter,
and sampling) on the signal

\begin{equation}
\tilde{r}_{k}(t)=\begin{cases}
r_{k-1}(t) & t\leq t_{k-1}\\
\tilde{r}(t) & t_{k-1}<t\leq t_{k}\\
0 & \text{else}
\end{cases}\,\,\,\,\text{with}\,\,\,r_{1}(t)=\tilde{r}(t)
\end{equation}
to detect $\hat{x}_{k}$. Importantly, \ac{DF-FNFT} requires to perform
at the \ac{RX} a total of one \ac{BNFT} and two \ac{FNFT}. Indeed,
as far as it concerns (i), at the $k$-th step one needs to evaluate
$r_{k-1}(t)$ performing a \ac{BNFT} only for $t\in(t_{k-2},t_{k-1}]$,
since the values for $t\leq t_{k-2}$ have already been evaluated
at the previous step, resulting overall in a single \ac{BNFT}. Regarding
(ii), similarly to the \ac{I-FNFT} case, one needs to add the contribution
of the signal in two symbol times for $t\in(t_{k-2},t_{k}]$, therefore
resulting in two \ac{FNFT}. 

Remarkably, both detection strategies, as well as \ac{DF-BNFT}, choose
the $k$-th symbol $x_{k}$ accounting only for its contribution in
the time window $(t_{k-1},t_{k}]$. While $x_{k}$ does not contribute
to the signal before $t_{k-1}$, it does for $t>t_{k}$, with this
contribution increasing at higher energies. Therefore, these detection
strategies do not consider all the available information, thus reducing
the effective \ac{SNR}. However, removing part of the signal also
improves performance, as shown in the next section. Moreover, for
what it concerns \ac{DF-FNFT}, considering $r_{k}(t)=\tilde{r}(t)$
for $t>t_{k-1}$ would drive to a much more computationally complex
detection: at the $k$-th step, one should perform \ac{FNFT} adding
the contribution of the signal to $N_{b}-k+2$ symbols (rather than
$1$).

\psection{System performance}

System performance was evaluated through simulations. The channel
is a standard single mode fiber of length $\mathcal{L}=\unit[4000]{km}$
(group velocity dispersion parameter $\beta_{2}=\unit[-20.39]{ps^{2}/km}$,
nonlinear coefficient $\gamma=\unit[1.22]{W^{-1}km^{-1}}$, and attenuation
$\alpha=\unit[0.2]{dB/km}$) with ideal distributed amplification
(spontaneous emission factor $\eta_{sp}=4$). The \ac{DAC} and \ac{ADC}
bandwidth is $\unit[100]{GHz}$. The symbol rate is $R_{s}=1/T_{s}=\unit[10]{GBaud}$
and the basic pulse $g(t)$ is Gaussian with $99$\% of the energy
contained into a symbol time $T_{s}$. To avoid overlapping of different
bursts during propagation and ensure the vanishing boundary conditions
of the \ac{NFT} type considered here, $N_{z}=160$ guard symbols
are inserted between different bursts. To account for the loss in
spectral efficiency due to guard symbols insertion between bursts
we consider the rate efficiency term $\eta=N_{b}/(N_{b}+N_{z})$ \cite{civelli2017noise}.

Numerical \ac{NFT} operations are performed with an oversampling
factor of $8$ samples per symbols. The \ac{FNFT} is numerically
performed using the Boffetta-Osborne method \cite{turitsyn2017optica,civelli2018OPEX},
while the \ac{BNFT} is computed with an enhanced version of the Nystrom
method \cite{civelli2018OPEX,civelliNFT}. System performance is measured
in terms of Q-factor as $Q_{\mathrm{dB}}^{2}=20\log_{10}[\sqrt{2}\mathrm{\text{erfc}}^{-1}(2P_{b})]$,
where the bit error rate $P_{b}$ is estimated by direct error counting.

The performance obtained through simulations are shown in Fig.\ \ref{fig:DF1-1}(a),
\ref{fig:DF1-1}(b), and \ref{fig:DF1-2}(a)  for $N_{b}=128,\,256,\,512$,
respectively. Firstly, the figures show that \ac{FNFT} standard detection
for \ac{NFDM} performs worse, as a consequence of being a detection
strategy not optimal in the nonlinear frequency domain. Secondly,
\ac{I-FNFT} performs better than \ac{FNFT} detection allowing for
an improvement of up to $\unit[3]{dB}$ without increasing the computational
complexity at all. Next, the figures show that a further performance
improvement can be achieved with \ac{DF-FNFT} detection, at the expense
of increasing the computational complexity by additionally performing
one \ac{FNFT} and one \ac{BNFT}. Finally, \ac{DF-BNFT} detection
provides the best performance, with a gain of up to about $\unit[3]{dB}$
with respect to \ac{DF-FNFT}, and about $\unit[7]{dB}$ with respect
to the conventional \ac{FNFT}.

Figure \ref{fig:DF1-2}(b) reports the optimal performance as a function
of the rate efficiency $\eta$. The figure shows that increasing the
rate efficiency, i.e., the number of information symbols per burst,
performance decreases \cite{civelli2018OPEX,Turitsyn_nature16}. Moreover,
Fig.\ \ref{fig:DF1-2}(a) emphasizes the relative behavior of the
considered detection strategies: \ac{FNFT} performs worse than all
others, \ac{I-FNFT} achieves better results than \ac{FNFT}, but
worse than \ac{DF-FNFT}, \ac{DF-BNFT} performs better than all others.

\begin{figure}
\hfill{}(a)\hfill{}\hfill{}(b)\hfill{}

\includegraphics[width=0.5\textwidth]{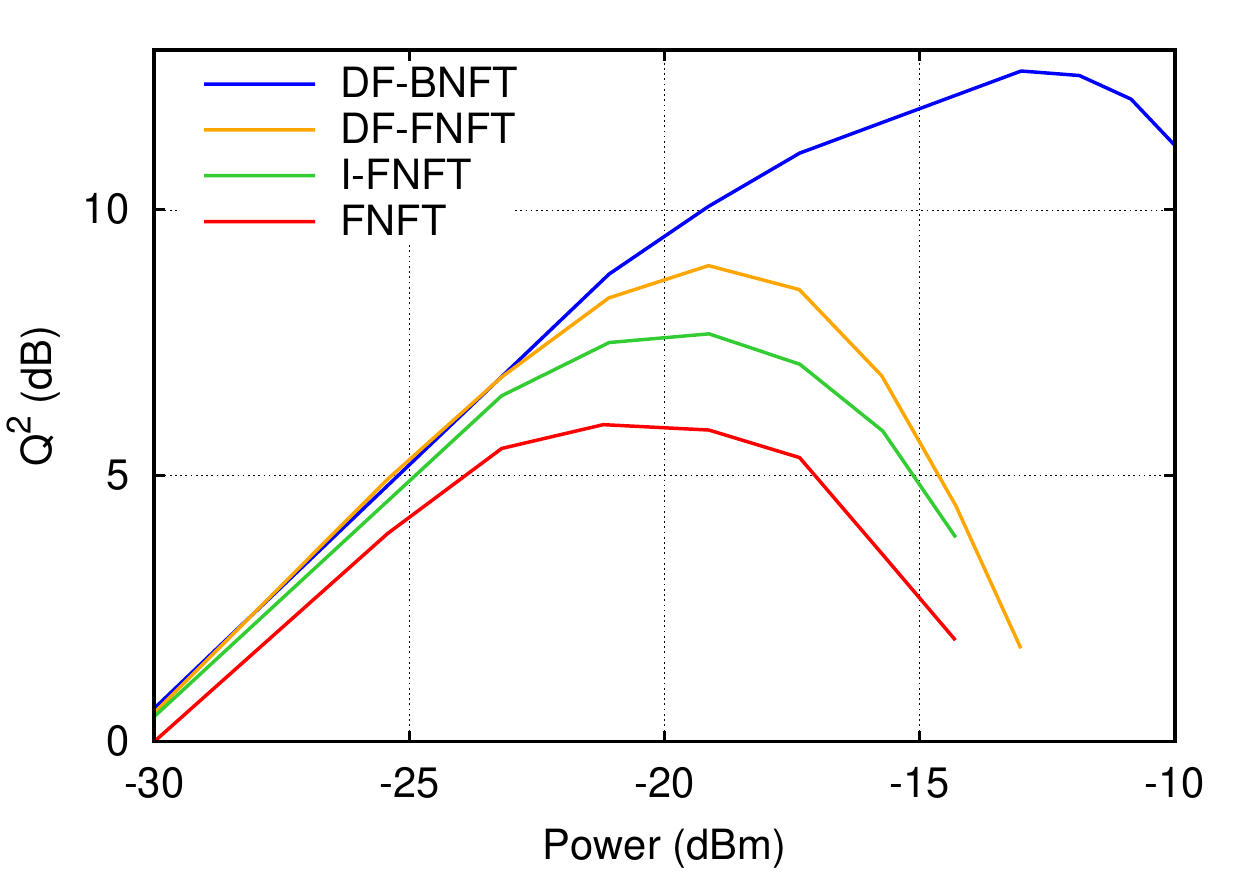}\includegraphics[width=0.5\textwidth]{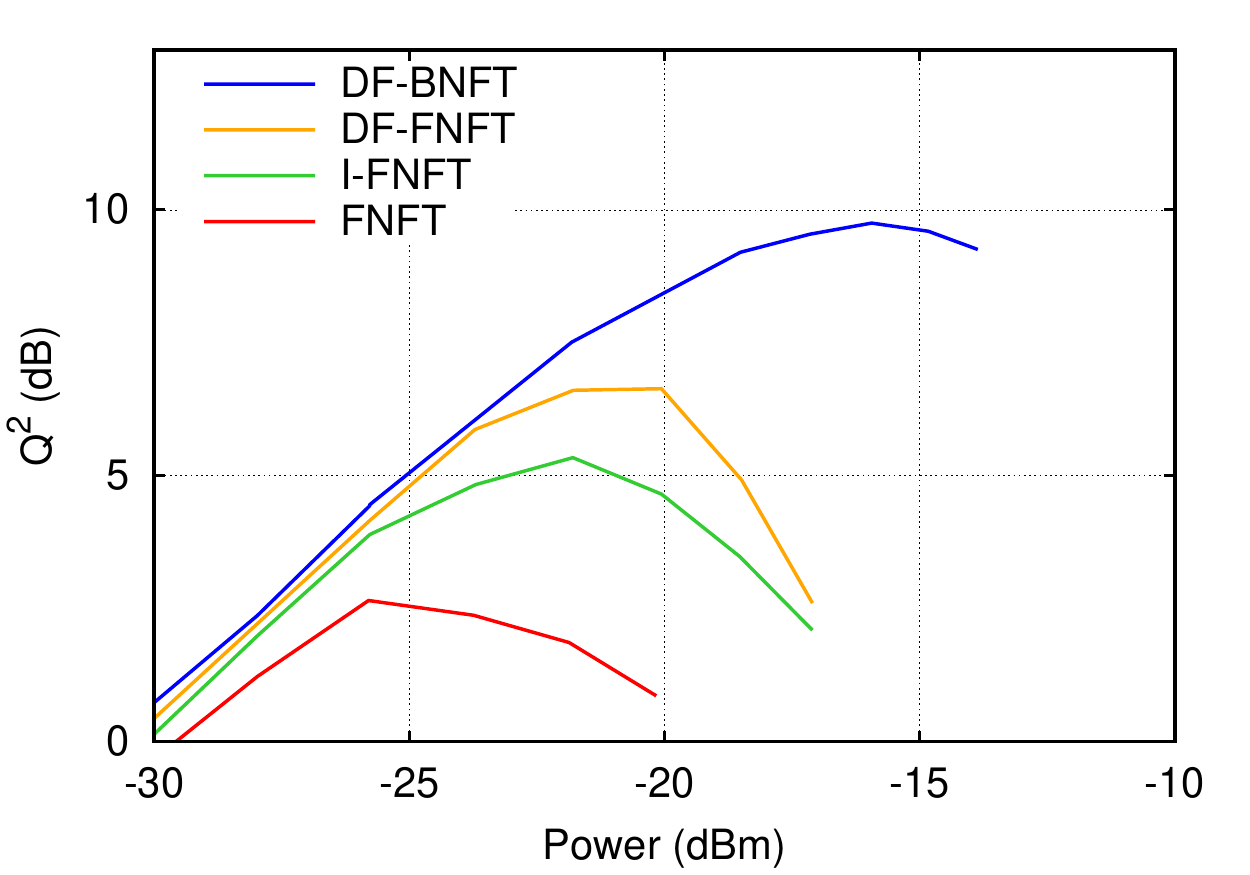}\caption{\label{fig:DF1-1}NFDM performance for different detection strategies
vs power for: (a) $N_{b}=128$ ($\eta=44\%$), and (b) $N_{b}=256$
($\eta=62\%$).}
\end{figure}
\begin{figure}
\hfill{}(a)\hfill{}\hfill{}(b)\hfill{}

\includegraphics[width=0.5\textwidth]{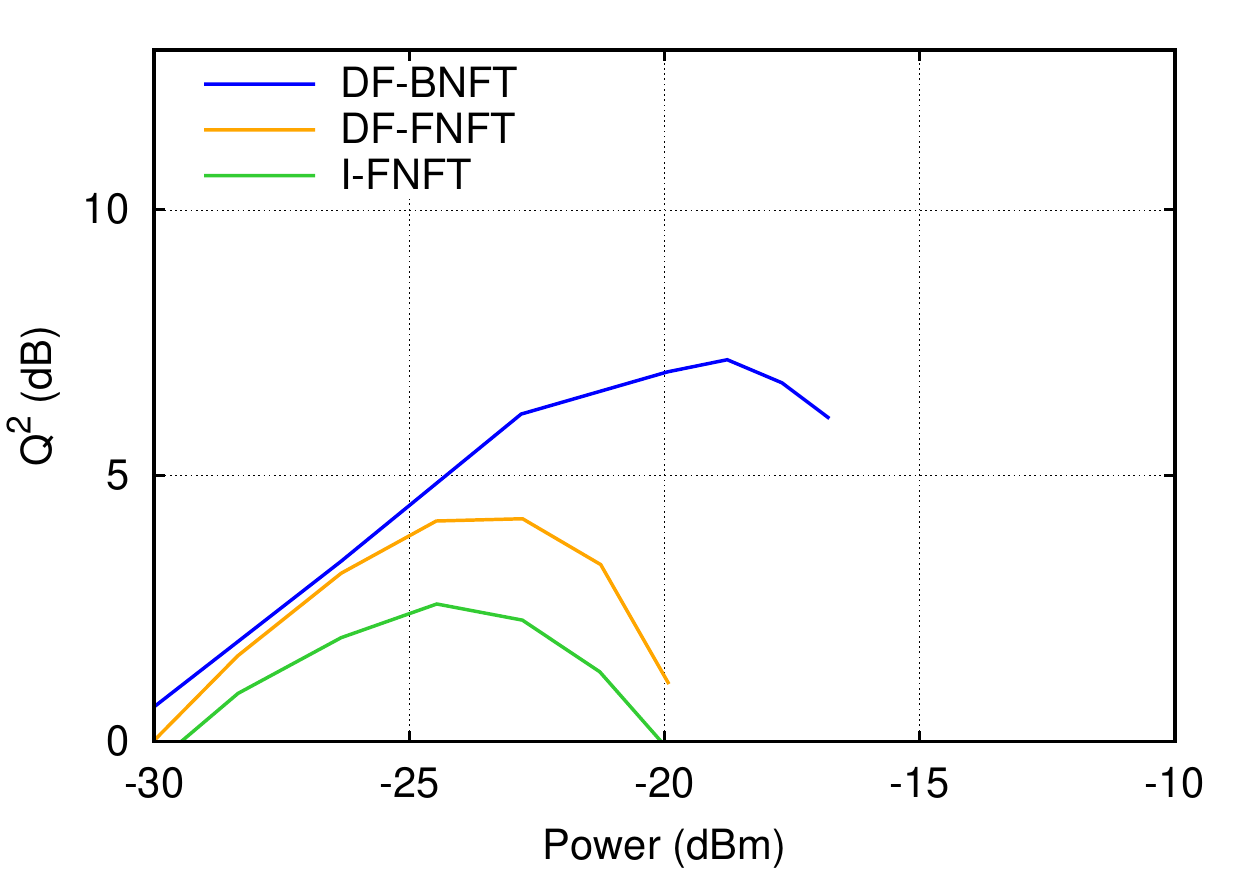}\raisebox{-2.5pt}{\includegraphics[width=0.48\textwidth]{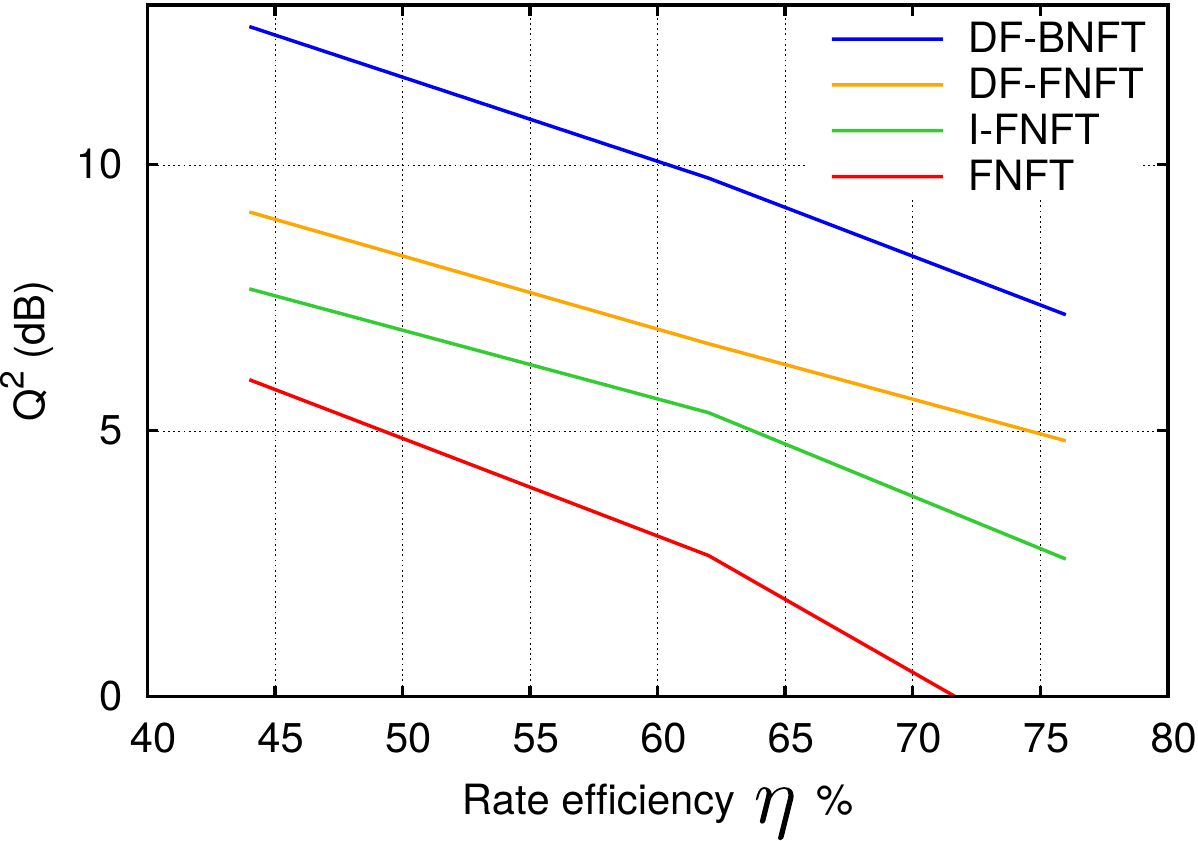}}\caption{\label{fig:DF1-2}NFDM performance for different detection strategies:
(a) vs power for $N_{b}=512$ ($\eta=76\%$), and (b) optimal performance
as a function of the rate efficiency $\eta$.}
\end{figure}

\psection{Conclusion}

In this paper, motivated by the fact that, performance-wise, the currently
considered detection strategy is one of the main critical aspects
of \ac{NFDM}, we proposed  two novel detection strategies specifically
designed for this transmission technique. We showed that, by relying
on an \ac{NFT} causality property,  the received signal can be partly
``cleaned'' before computing its nonlinear spectrum, thus reducing
the detrimental ``noise amplification'' effect taking place on the
latter and badly affecting conventional \ac{FNFT} detection. In the
\ac{I-FNFT} detection technique, the noise cleaning effect is limited
to the signal portion following the symbol to be detected; in the
\ac{DF-FNFT} technique, it is extended also to the previous part
of the signal by using decision feedback. In both cases, decisions
are made in the nonlinear frequency domain, after matched filtering
and symbol-time sampling, exactly as in conventional \ac{FNFT} detection
and in contrast to \ac{DF-BNFT} detection, which completely avoids
the noise amplification effect by making decisions in the time domain
\cite{civelli2018OPEX}. We compared the two novel detection strategies
with the conventional \ac{FNFT} and the \ac{DF-BNFT} ones. Both
the \ac{I-FNFT} and \ac{DF-FNFT} strategies perform better than
the conventional \ac{FNFT} one, with gains of up to 3 and 5~dB,
respectively, proving that currently considered \ac{NFDM} schemes
are not optimized and can be enhanced. Even if both detection strategies
turn out to be inferior to the \ac{DF-BNFT} strategy, their computational
complexity is considerably lower and comparable to that of the conventional \ac{FNFT}
one.\bibliographystyle{IEEEtran}

\end{paper} 
\end{document}